\documentstyle[fleqn,twoside,psfig]{article}
\begin{document}

\noindent{\huge THE SEARCH FOR THE QUARK-GLUON PLASMA}
\bigskip\bigskip

\noindent{\large {\it John W. Harris}}
\medskip

\noindent Lawrence Berkeley Laboratory, University of California, 
Berkeley, CA 94720 
\medskip

\noindent {\large {\it Berndt M\"uller}}
\medskip

\noindent Department of Physics, Duke University, Durham NC 27708-0305
\bigskip

\noindent {\sc key words}: QCD, deconfinement, chiral symmetry restoration,
high density nuclear matter, relativistic heavy ion collisions, AGS, SPS, 
RHIC, LHC, quark matter, phase transitions
\bigskip
\hrule
\bigskip

\begin{abstract}

We provide an overview of the present understanding of the transition from
hadrons to a quark-gluon plasma, its signatures, and the experimental
results so far. We discuss results of numerical simulations of the 
lattice gauge theory and critically evaluate the various observables 
that have been proposed as signatures of the QCD phase transition. 
We place the existing data from relativistic heavy ion experiments at 
the Brookhaven AGS and CERN SPS into perspective and provide an overview 
of the techniques and strategies that will be employed in the search for 
the quark-gluon plasma at heavy ion colliders, such as RHIC and the LHC. 

\end{abstract}

\bigskip
\hrule
\bigskip

\tableofcontents

\bigskip

\section{\rm INTRODUCTION}

\subsection{\it The QCD phase transition}

Strongly interacting matter is described at the fundamental level through
the interaction of quarks by the exchange of gluons.  This non-abelian
gauge theory, called quantum chromodynamics (QCD), exhibits a number of 
remarkable features.  (1) At short distances or large momenta $q$, the 
effective coupling constant $\alpha_s(q^2)$ decreases logarithmically, 
i.e. quarks and gluons appear to be weakly coupled.  (2) At large distances 
or small momenta, the effective coupling becomes strong, resulting in 
the phenomena of quark confinement\footnote{Quark confinement is the 
technical term describing the observation that quarks do not occur 
isolated in nature, but only in hadronic bound states as mesons and baryons.}
and chiral symmetry breaking\footnote{Chiral symmetry breaking expresses 
the fact that quarks confined in hadrons do not appear as nearly massless 
constituents but are endowed with a dynamically generated mass of several 
hundred MeV.}. (3) At low energies, the QCD vacuum is characterized by 
nonvanishing expectation values of certain operators, usually called 
vacuum condensates, which characterize the nonperturbative physical 
properties of the QCD vacuum.  Most important for our discussion are
\cite{Shi79}:
\begin{itemize}
\item the quark condensate $\langle \bar\psi\psi\rangle \approx$ (235 
MeV)$^3$ and,

\item the gluon condensate $\langle \alpha_s G_{\mu\nu}G^{\mu\nu}\rangle
\approx$ (500 MeV)$^4$.
\end{itemize}
The quark condensate describes the density of quark-antiquark pairs
found in the QCD vacuum and is the expression of chiral symmetry breaking.
The gluon condensate measures the density of gluon pairs 
in the QCD vacuum and is a manifestation of the breaking of scale
invariance of QCD by quantum effects.

It is not uncommon in nature that spontaneously broken symmetries are
restored at high temperature through phase transitions.  Well known 
examples are ferromagnetism, superconductivity, and the transition from
solid to liquid.  More closely connected to our subject,
nuclear matter at low temperatures has a dense liquid phase,
which goes over into a dilute gaseous phase at $T>5$ MeV.  Evidence for 
this phase transition has recently been seen in nuclear collisions at 
intermediate energies \cite{Poc95}.

As the temperature increases in QCD, the interactions among 
quanta occur at ever shorter distances, governed by weak coupling, 
while the long-range interactions become dynamically screened.
This picture is supported by finite temperature
perturbation theory, showing that the effective coupling constant
$\alpha_s(T)$ falls logarithmically with increasing temperature
\cite{Col75} and also by more general arguments \cite{Pol77}.  As a
consequence, nuclear matter at very high temperature exhibits neither 
confinement nor chiral symmetry breaking.  This new phase of QCD is 
called the quark-gluon plasma.

Since there exist order parameters, such as the quark condensate, 
which vanish at high temperature,\footnote{Two technical notes:
(1) this statement neglects the effect of finite current quark masses;
(2) the situation is somewhat unusual for the deconfinement transition in
the pure non-abelian gauge theory.  Here the Z(3) center symmetry is
broken spontaneously at {\it high} temperature, allowing free quarks
to exist, whereas it is manifest in the QCD vacuum, causing quark
confinement \cite{Pol95}.} 
there are good reasons to expect that the transition between the
low-temperature and high-temperature manifestations of QCD is not
smooth but exhibits a discontinuity, i.e. a phase transition.
The order of the chiral phase transition is believed to be quite
sensitive to the number of light, dynamical quark flavors.
Universality arguments \cite{Pis84,Raj93a} predict a second-order phase
transition for two massless flavors and a first-order transition for
three massless flavors.  Numerical simulations of the lattice gauge 
theory (see section 2.1) have established the transition temperature to
lie in the range $150 \pm 10$ MeV at vanishing net quark density.

According to the standard cosmological model \cite{Kol90}, the temperature 
of the cosmic background radiation exceeded 200 MeV during the first 10 
$\mu s$ after the Big Bang.  The early universe was hence filled with a
quark-gluon plasma, rather than hadrons.  Hence,
physical processes occurring during this very early
period can be described in terms of quark and gluon transition
amplitudes rather than hadronic amplitudes. This facilitates reliable
calculations of transport processes, such as baryon number violating
processes during the electroweak phase transition.

Chiral symmetry is also expected to be restored at high baryon
density even at zero temperature.  Many model studies of this
phenomenon have been performed, yielding critical densities $4\rho_0 <
\rho_c < 10\rho_0$, where $\rho_0$ denotes the ground state density
of nuclear matter.  Since ab-initio calculations based on lattice
QCD are not feasible yet, the uncertainty of $\rho_c$ remains large.  One
expects a smooth connection between the high-$T$ and high-$\rho$ phase
transitions, giving rise to a continuous phase boundary $T_c(\rho)$.
For $T<T_c(\rho)$ the effective description of
strongly interacting matter at low momenta is in terms of hadronic
degrees of freedom (baryons and mesons), whereas for $T>T_c(\rho)$ the
effective degrees of freedom at low momenta carry the quantum numbers
of quarks and gluons.

It is important to recognize that this difference is only discernible
at momenta below the chiral symmetry breaking scale $q_{\rm CSB}^2
\approx (4\pi f_{\pi})^2 \approx $ 1 GeV$^2$. For processes involving
momentum transfers above that scale light quarks are effectively
massless and deconfined in either phase.  An effective description 
above $q_{\rm CSB}^2$ must, therefore, always be based on the
elementary degrees of freedom (quarks and gluons).  Hence,
experimental signatures for the change in the structure of strongly
interacting matter must be sensitive to the dynamics of the low-energy
degrees of freedom of QCD.  Quarks and gluons are already known to
provide the correct description for momenta $q^2\gg q^2_{\rm CSB}$.
Quark-gluon plasma signals must probe the momentum range $q^2<
q_{\rm CSB}^2$ and show that current quarks and gluons remain effective
degrees of freedom also in this range when $T>T_c(\rho)$.

\subsection{\it Abnormal nuclear matter}

Many speculations of additional phases of baryon-rich nuclear
matter exist:  pion condensates \cite{Mig71,Bro76}, density isomers
\cite{Lee74}, nuclear matter with a large strangeness content
\cite{Bod71,Chi79,Wit84,Far84,Kap86}, and others.  Of these, kaon 
condensates and strange quark matter appear to be the best established.  
In some models of hadron structure it is not even possible to exclude that 
strange quark matter is the true ground state of nuclear matter 
\cite{Kas93,Gil93,Mad93,Des93}.  In other models, strange quark 
matter may be unstable against weak decay, but stable under the strong 
interactions.  The basic argument for an increased stability of strange 
matter is that the Fermi energy can be lowered significantly by 
distributing the given baryon number of three quark flavors rather than 
only two.  The price to be paid is the higher current quark mass of the 
strange quark and, possibly, the destruction of the color singlet 
structure of baryons.  The energy balance for these different aspects is 
delicate, hence reliable predictions are impossible without a
better knowledge of baryon structure in the framework of QCD.

Mechanisms for the separation of strange quark matter have been
conceived in the environment of the early universe \cite{Wit84} as 
well as in relativistic heavy ion reactions \cite{Liu84,Gre87,Gre91,Cra92}.  
While there exist rather strong astrophysical limits on the density of 
strange quark matter ``nug\-gets'' (strangelets) in our universe
\cite{Mad88}, the search for strange quark matter produced in heavy ion 
collisions has only just begun.  The experimental signature for light 
multi-strange nuclei would be an abnormally low charge-to-mass ratio.

\section{\rm THEORETICAL GUIDANCE AND EXPECTATIONS}

\subsection{\it Lattice gauge theory}

Lattice gauge theory \cite{Wil74,Cre83} allows for a potentially exact,
nonperturbative numerical calculation of observables in QCD.
Improvements of the original algorithms,  together with 
significant increases in computing power due to parallel processing, 
have permitted fairly reliable evaluations of the thermodynamic 
averages of many interesting quantities, which can be extrapolated 
to the infinite volume limit.  State-of-the-art simulations of the 
finite-temperature lattice gauge theory employ lattices with spatial 
size $12^3$ and 4 points in the Euclidean time direction \cite{Blu95}.
Simulations of pure SU(3) gauge theory without quarks are possible 
on much larger lattices, such as $32^3\times 12$ \cite{Boy95}.

The simulations without dynamical quarks clearly exhibit a first-order 
phase transition at a temperature of $T_c \approx $\quad 260 MeV.  
Below $T_c$ the free energy of an isolated quark is infinite, above
$T_c$ it is finite.  Hence quarks are not confined in the 
high-temperature phase.  It is also possible to evaluate the quark 
condensate $\langle \bar\psi\psi\rangle$ in the so-called ``quenched'' 
approximation, where the influence of quark-antiquark pairs on the QCD 
vacuum is neglected.  This quantity shows a strong drop over the same 
temperature range (see Figure 1), indicating that deconfinement and 
restoration of chiral symmetry go hand-in-hand.

Simulations of lattice QCD with dynamical quarks have not
yet overcome the limitations due to finite lattice size.  Present
results indicate a smooth cross-over between phases for $N_f=2$ light
quark flavors and a first-order phase transition for $N_f\ge 3$.
Figure 2 shows recent results (at $N_f=2$) for the energy density 
$\epsilon$ and pressure $P$ as a function of temperature.  $\epsilon 
(T)/T^4$ shows a dramatic rise.  The thermal change in the pressure $P$ 
is much smoother, causing a large difference between $3P$ and $\epsilon$
in the transition region.  Since $(3P-\epsilon)$ is a measure of
nonperturbative interactions, this shows that interactions continue to
be influential and that the high-temperature phase is not simply a
gas of quasi-free quarks and gluons in this temperature range.  The 
effect can be successfully described assuming that the propagating 
degrees of freedom have an effective thermal mass \cite{Bir90,Pes94}

\subsection{\it Thermal perturbation theory}

Many of the properties of the quark-gluon plasma far above $T_c$ can 
be calculated in the framework of thermal perturbation theory.  Neglecting 
current quark masses, the equation of state up to order $g^2$ is given by 
\cite{Bay76,Fri78,Shu80}
\begin{eqnarray}
\epsilon &= &\left( 1-{15\over 16\pi^2} g^2 \right) {8\pi^2\over15} T^4 +
N_f \left( 1 - {50\over 84\pi^2} g^2\right) {7\pi^2\over 10} T^4 +
\nonumber \\
&&\qquad \sum_f \left( 1 - {2\over 4\pi^2} g^2\right) {3\over \pi^2}\;
\mu_f^2 \left( \pi^2 T^2+{1\over 2}\mu_f^2\right).
\end{eqnarray}
where $f$ denotes the quark flavors and $\mu_f$ the quark chemical
potential of each flavor.  Higher order corrections have been
calculated for $\mu_f=0$, up to order $g^5$ \cite{Arn94}.  Various
arguments can be made that $\alpha_s$ should be taken at an effective
momentum scale of the order $2\pi T$, corresponding to $\alpha_s
\approx 0.3$ at the critical temperature, or $g\approx 2$.

Additional insight into the properties of the interacting quark-gluon
plasma is obtained by considering the dispersion relations of small
perturbations carrying the quantum numbers of quarks or gluons
\cite{Kli82,Wel82}.  These excitations govern the dissipation mechanisms
in the quark-gluon plasma.  For gluonic excitations, one obtains two 
different modes, with transverse and longitudinal polarization, 
respectively.  Soft gluonic excitations, also called plasmons, carry an 
effective mass of order $gT/\sqrt{3}$.  The plasmon mode is strongly damped.  
In the limit $k\to 0$ the plasmon decay width is $\Gamma\approx {1\over 2}
g^2T$, which provides an estimate for the thermalization time of a
quark-gluon plasma \cite{Bir95}.  The static longitudinal gluon propagator 
is screened with screening mass $m_E\approx gT$, whereas the transverse 
propagator remains unscreened (lack of magnetic screening).  Lattice 
simulations indicate a localization of static color-magnetic fields at 
the momentum scale $g^2T$.  The screening of color-electric fields lies
at the origin of the deconfinement of quarks at high temperature.

\subsection{\it Dynamical models of relativistic heavy ion collisions}

\subsubsection*{\sc Parton Cascades}

Relativistic heavy ion collisions are the most promising tool for creating
a quark-gluon plasma in the laboratory.  QCD predicts that energy density
at midrapidity grows like $A^{2/3}$, where $A$ is the nuclear mass
\cite{Hwa86,Bla87a}, but at most logarithmically with the center-of-mass
energy.  In order to reach temperatures far above $T_c$, the initial 
kinetic energy of the nuclei must be rapidly thermalized on a time scale 
of order 1 fm/$c$.  Early ideas about the mechanism of energy deposition 
were based either on the inside-outside cascade model of parton scattering 
\cite{Ani80} or the breaking of color flux tubes \cite{Bia85,Kaj85}. 
More recently, detailed microscopic models have been constructed 
\cite{Wan91,Gei92,Gei95} that permit the study of the energy deposition 
process in space-time as well as in momentum space, in the framework of 
perturbative QCD.  The models are based on the concept that the colliding 
nuclei can be decomposed into their parton substructure.
The perturbative interactions among these partons 
can then be followed until thermalization.  One finds that partonic 
cascades account for at least half the expected energy deposition at RHIC 
and an even larger fraction in the energy range of the LHC \cite{Esk94}.

Parton cascade models predict a very rapid thermalization of the
deposited energy.  This is caused by a combination of radiative energy
degradation and separation of partons with widely different rapidities
due to free-streaming.  The transverse momentum distribution of initially
scattered partons is already to a high degree exponential if radiative
processes are taken into account.  Free-streaming causes the local 
longitudinal momentum distribution of partons to coincide with the 
transverse distribution after a time approximately equal to the mean time
between parton interactions.  The models predict that thermalization occurs 
on a proper time scale of 0.3--0.5 fm/$c$ at RHIC energies\cite{Esk94}.

Due to the large cross sections and higher branching probabilities 
of gluons, the thermalized parton plasma is initially gluon rich and 
rather depleted of quarks \cite{Shu92}.  Chemical equilibration of the
parton plasma proceeds over a time of several fm/$c$ in most scenarios
\cite{Bir93,Gei93a}, but may be faster if higher-order QCD processes are
important \cite{Xio94}.

Another interesting issue concerns the homogeneity of initial conditions.  
Partonic cascades can lead to a rather uneven energy deposition, because 
of cross section fluctuations.  ``Hot spots'', caused by strongly 
inelastic parton scatterings could lead to observable, nonstatistical 
fluctuations in the final hadron distribution \cite{Gyu95a}.

\subsubsection*{\sc Hydrodynamic Models}

After (local) momentum equilibration further evolution of the
quark-gluon plasma to its final dissolution can be described in the
framework of relativistic hydrodynamics.  According to the results of
parton cascade models, the initial conditions for this evolution in
the central rapidity region are boost invariant to a large degree, as
anticipated by Bjorken \cite{Bjo83}.  Assuming purely longitudinal
expansion, the temperature then falls as $\tau^{-1/3}$, where $\tau$
is the local proper time.  Cooling is substantially enhanced by the 
transverse expansion that is generated by the high internal pressure of the 
plasma, when the initial temperature is significantly above $T_c$.  Typical 
estimates of the plasma lifetime are 4 fm/$c$, after which a mixed 
quark-hadron phase is formed in a first-order phase transition
\cite{Bla91a}.  Due to transverse expansion, however, even the mixed
phase decays on a time scale of 10 fm/$c$.  The lifetime of the mixed
phase could be longer, if the plasma were formed at the
critical temperature, where the pressure is minimal \cite{Hun95,Ris95}.  
This is expected to occur at energies far below those accessible at RHIC 
-- likely between the current AGS and SPS energies -- in a regime where our 
understanding of the thermalization mechanism is rather limited.  
A long-lived $(\gg 10$ fm/$c$) mixed phase could be detected by its effect
on two-particle correlations \cite{Pra84,Ber88}.

The hydrodynamic approach becomes invalid when the typical distance
between particles exceeds the mean free path.  This happens rather
shortly after the quark-hadron phase transition, when the temperature
falls below $120-130$ MeV \cite{Gav92,Hag94a}.  Since various hadrons have
different mean free paths, the freeze-out is differential with K$^+$-mesons
freezing out first, followed by nucleons, K$^-$ and, finally, pions.

\section{\rm PLASMA SIGNATURES}

Experimental investigations of the quark-gluon plasma require the
identification of appropriate experimental tools for observing its 
formation and studying its properties.  One serious problem is that 
the size and lifetime of the plasma are expected to be small, at most a 
few fermi in diameter and perhaps 5 to 10 fm/$c$ in duration.  
Furthermore, signals of the quark-gluon plasma compete with
backgrounds emitted from the hot hadronic gas phase that follows the 
hadronization of the plasma, and are modified by final state
interactions in the hadronic phase.  In spite of this, a wealth of ideas 
has been proposed in the past decade as to how the identification 
and investigation of the short-lived quark-gluon plasma phase could be 
accomplished.  It is beyond the scope of this review to present a 
comprehensive survey of quark-gluon plasma signatures. We will therefore
concentrate on the most promising ones.  More details can be found 
elsewhere \cite{Kaj87,Ber90,Sin93,Mul95}.

\subsection{\it Kinematic probes}

The basic concept behind this class of signatures is the determination
of the energy density $\epsilon$, pressure $P$, and entropy density $s$ 
of superdense hadronic matter as a function of the temperature $T$ and the 
baryochemical potential $\mu_B$.  One seeks to observe a rapid 
rise in the effective number of degrees of freedom, as expressed by the 
ratios $\epsilon/T^4$ or $s/T^3$, over a small temperature range.  

Measurable observables that are related to the variables $T$, $s$, and
$\epsilon$, are customarily identified with the average transverse momentum
$\langle p_{\rm T}\rangle$, the hadron rapidity distribution $dN/dy$, and
the transverse energy $dE_{\rm T}/dy$, respectively \cite{Hov82}.  One can, 
in principle, invert the $\epsilon$-$T$ diagram of Figure 2 and plot 
$\langle p_{\rm T}\rangle$ as a function of $dN/dy$ or $dE_{\rm T}/dy$.  
If a rapid change in the effective number of degrees of freedom occurs, 
one expects an S-shaped curve, whose essential characteristic feature is
the saturation of $\langle p_{\rm T}\rangle$ during the persistence of a
mixed phase, continuing into a second rise when the structural
change from color-singlet to colored constituents has been completed.

The high pressure of the quark-gluon plasma leads to the formation of a
collective outward flow during the expansion of the dense matter.
Detailed numerical studies in the context of the hydrodynamical model 
have shown that this characteristic transverse flow of particles is rather 
weak in realistic models \cite{Ger86,Kat92}.  The strength of this signal
can be enhanced by studying higher moments of the momentum distribution, 
or heavier hadrons such as baryons \cite{Bla86a}.  The transverse flow 
signal would be enhanced by the formation of a detonation wave during
the hadronization transition \cite{Gyu84,Bar85,Sei87,Bil93}.

Identical particle interferometry, e.g. $\pi\pi$, KK, or NN correlations,
yields information on the reaction geometry and provides important
information about the space-time dynamics of nuclear collisions.  By 
studying the two-particle correlation function along
various directions in phase space, it is possible to obtain measurements 
of the transverse and longitudinal size, of the lifetime, and of flow
patterns of the hadronic fireball at the moment when it breaks up
into separate hadrons \cite{Pra84,Ber88}.  Recent theoretical work has 
shown the importance of the finite lifetime of the fireball \cite{Cso94}, 
of flow patterns \cite{Cso95}, and of shadowing effects \cite{Chu94}.
Since interferometric size determinations will be possible on an
event-by-event basis for collisions of heavy nuclei at the SPS, RHIC,
and LHC, the correlation of global parameters like 
$\langle p_{\rm T}\rangle$ and $dN/dy$ with the fireball geometry can
be performed on individual collision events.

\subsection{\it Electromagnetic probes}

Photons and lepton pairs provide probes of the interior of the 
quark-gluon plasma during the earliest and hottest phase of the evolution 
of the fireball since they are not affected by final state interactions.  
Unfortunately, these probes have rather small yields and must compete with 
relatively large backgrounds from hadronic processes, especially 
electromagnetic hadron decays.  

In the hadronic phase, the electromagnetic response function is dominated 
by the $\rho^0$ resonance at 770 MeV.  On the other hand, perturbative QCD 
predicts a broad continuous spectrum above twice the thermal quark mass
$m_q=gT/\sqrt{6}$ in the high temperature phase.  Below 100 MeV collective 
modes are predicted to exist in both phases.  

\subsubsection*{\sc Lepton Pairs} 

Many of the original calculations on lepton pairs as probes of the 
quark-gluon plasma \cite{Fei76,Shu78,Dom81,Kaj81,Chi82,Hwa85,Cle86,Kaj86}
concentrated on invariant masses in the range below the $\rho$-meson mass.
With an improved understanding of the collision dynamics and
the hadronic backgrounds \cite{Cle91,Gal94}, it has since become clear 
\cite{Ruu91} that lepton pairs from the quark-gluon plasma 
can probably only be identified for invariant masses above $1-1.5$ GeV.
At the high-mass end, the yield of Drell-Yan pairs from first 
nucleon-nucleon collisions exceeds the thermal dilepton yield.

Recent progress in understanding the mechanisms of thermalization has 
revealed that the yield of high-mass dileptons critically depends on, and 
provides a measure of, the thermalization time \cite{Kap92}.  Lepton pairs 
from the equilibrating quark-gluon plasma may dominate over the Drell-Yan 
background up to masses in the range $5-10$ GeV, as predicted by the parton 
cascade \cite{Gei93b} and other models of the early equilibration phase of 
the nuclear collision \cite{Shu93,Kam92}.  If this turns out to
be true, the early thermal evolution of the quark-gluon phase can be traced in 
a rather model independent way \cite{Str94}.  Dileptons from 
charm decay are predicted to yield a substantial contribution to the total 
dilepton spectrum and could, because of their different kinematics, provide 
a measure of the total charm yield \cite{Vog93}, which may be enhanced
due to rescattering of gluonic partons \cite{Lin94,Lev95}, if the
direct background \cite{Sar94} is sufficiently well understood.

Lepton pairs from hadronic sources in the invariant mass range
between 0.5 and 1 GeV are important signals of the dense hadronic
matter formed in nuclear collisions \cite{Sie85,Sei92}.  They provide
exclusive information about possible medium modifications of hadronic
properties, especially of the $\rho$-meson, at high density \cite{Gal87,Kar93}.
Another strategy for using the leptonic $\rho$-meson decay as a probe of
the hadronic phase of the fireball is based on the idea that the 
$\rho$-peak is expected to grow strongly relative to the $\omega$ 
peak in the lepton pair mass spectrum, if the fireball lives 
substantially longer than 2 fm/$c$.  Because of the short average lifetime 
of the $\rho$-meson, the $\rho/\omega$ ratio can therefore serve as a 
fast ``clock'' for the fireball lifetime \cite{Hei91}.

\subsubsection*{\sc Direct Photons}  

In contrast to the lepton-pair spectrum the hadronic radiation
spectrum is not concentrated in a single narrow resonance.  The dominant
source of photons from the thermal hadron gas is the 
$\pi\rho\to\gamma\rho$ reaction \cite{Kap91}, to which the broad $a_1$ 
resonance may be an important contribution \cite{Xio92}.  In the quark phase 
the scattering process $gq\to\gamma q$ dominates.  Infrared
singularities occurring in perturbation theory are softened by screening
effects \cite{Kap91,Bai92}.  The result is that a hadron gas and a 
quark-gluon plasma in the vicinity of the critical temperature $T_c$ emit 
photon spectra of roughly equal intensity and similar spectral shape.

However, a clear signal of photons from the quark-gluon plasma could be 
visible for transverse momenta $p_T$ in the range $2-5$ GeV/$c$ if a very hot 
plasma is formed initially \cite{Str94,Sri92,Cha92}.  The photon spectrum 
in the $p_{\rm T}$ range $1-2$ GeV/$c$ is mostly emitted from the mixed 
phase.  Transverse flow effects make the separation of the 
contributions from the different phases more difficult \cite{Ala93}, and 
destroy the correlation between the slope of the photon spectrum in the 
intermediate $p_{\rm T}$ range and the temperature of the mixed phase 
\cite{Neu95}.

\subsection{\it Probes of deconfinement}

\subsubsection*{\sc Quarkonium Suppression}

The suppression of $J/\psi$ production \cite{Mat86} in a quark-gluon plasma
occurs because a $c\bar c$ pair formed by fusion of two gluons from the 
colliding nuclei cannot bind inside the quark-gluon plasma \cite{Meh88}.  
Lattice simulations of SU(3) gauge theory \cite{Deg86,Kan86} show that this 
condition should be satisfied already slightly above the deconfinement 
temperature.  The screening length appears to be even shorter when 
dynamical fermions are included in the lattice simulations
\cite{Kar88a,Bla91b}.  Excited states of the $(c\bar c)$ system, such as 
$\psi'$ and $\chi_c$, are more easily dissociated and should disappear 
as soon as the temperature exceeds $T_c$.  For the heavier $\Upsilon 
(b\bar b)$ system similar considerations apply, although shorter screening 
lengths are required than for the charmonium states \cite{Kar91}.  The 
dissociation temperature of the $\Upsilon$ ground state is predicted to be 
around 2.5 $T_c$, that of the larger $\Upsilon'$ state only slightly
above $T_c$.

Owing to its finite size, the formation of a $(c\bar c)$ bound state
requires a time of the order of 1 fm/$c$ \cite{Cer90,Huf90,The90}.  The 
$J/\psi$ may still survive, if it escapes from the region of high density 
and temperature before the $c\bar c$ pair has been spatially separated by 
more than the size of the bound state \cite{Mat86}.  This will happen either 
if the quark-gluon plasma cools very fast, or if the $J/\psi$ has 
sufficiently high transverse momentum 
\cite{Bla87b,Kar88b,Gaz91,Lie91}.  On the other hand, the charmonium may 
also be destroyed by sufficiently energetic collisions with comoving hadrons, 
leading to dissociation into a pair of $D$-mesons \cite{Gav88a,Vog88}.  
Dissociation via quark exchange with mesons composed of light quarks, such 
as the $\rho$-meson, has been estimated in a nonrelativistic quark model 
\cite{Mar95a} to reach several mb.  Similar values are obtained, if 
$J/\psi$ production is fed by a large fraction of easily absorbed 
color-octet $(c\bar c)$ states \cite{Kha95}.  Additional effects that can 
contribute to $J/\psi$ suppression even in hadron nucleus interactions are 
nuclear shadowing of soft gluons, initial state scattering of partons 
resulting in a widened transverse momentum distribution, and final state 
absorption on nucleons \cite{Gup92,Vog92,Don93}.  Suppression mechanisms
based on interactions with comoving particles generally predict that the
$\psi'$ state should be more strongly suppressed than the $J/\psi$ 
\cite{Kar91,Gav93a}.  This holds equally for a quark-gluon plasma as
for a comoving thermalized gas of hadrons.

\subsubsection*{\sc Strangeness Enhancement}

The production of hadrons containing strange quarks is normally suppressed
in hadronic reactions compared with the production of hadrons containing
only up and down valence quarks \cite{Bai87}.  This suppression increases 
with growing strangeness content of the produced hadrons.  
When a quark-gluon plasma
is formed, the production of hadrons carrying strange quarks is expected
to be saturated because the strange quark content of the plasma is
rapidly equilibrated by $s\bar{s}$ pair production in interactions of
two gluons \cite{Raf82a}.  As a result, the yield of multi-strange baryons
and strange antibaryons is predicted to be strongly enhanced in the
presence of a quark-gluon plasma \cite{Raf82b,Koc86}.

In addition, the relative abundances of the various strange particle 
species (mesons, strange and multi-strange baryons, and their antiparticles)
allow the determination of relative strangeness equilibrium, saturation in 
the overall strangeness content ($\gamma_{s}$), and strangeness neutrality 
in a thermo-chemical approach \cite{Hei94}. These ratios can be 
calculated assuming either a hadron gas scenario or a quark-gluon plasma 
scenario, and a comparison can be made of the values extracted from 
the models in the two scenarios in conjunction with other thermodynamic
variables of the system, such as the temperature $T$, the baryo-chemical 
potential $\mu_{B}$, and the entropy \cite{Let92,Let94a}.

Because strange hadrons interact strongly, their final state interactions
have to be modeled in considerable detail, before firm predictions about 
strange hadron yields are possible.  Theoretical studies \cite{Koc86,Bar88}
have shown that an enhanced strangeness content cannot be destroyed nor
generated by interactions during the break-up phase.  Fragmentation
processes during the hadronization phase transition can contribute 
significantly to the final abundances of strange hadrons \cite{Bar93a}, 
but this does not invalidate the usefulness of strangeness enhancement
as a plasma signature.

\subsection{\it Probes of chiral symmetry restoration}

\subsubsection*{\sc Disoriented Chiral Condensates}
\medskip

The temporary restoration of chiral symmetry in nuclear collisions may 
result in the formation of domains of disoriented chiral condensate (DCC) 
\cite{Bjo92}.  This term describes a coherent excitation of the pion field
corresponding to a local misalignment of the chiral order parameter $\langle
\bar\psi\psi\rangle$.  Such domains would decay into neutral and charged 
pions, favoring pion charge ratios $N_{\pi^0}/N_{\pi}$ substantially 
different from one-third.  This could explain why final states with a large
fraction of charged pions over neutral pions, observed in Centauro events 
\cite{Lat80}, can occur with significant probability \cite{Lam84,Pra93}.  

A DCC can be described as a nonlinear wave in the sigma model 
\cite{Ans89,Bla92}.  Such a wave can be excited by the growth of local 
instabilities during the transition from the chirally restored 
high-temperature phase of QCD to the low-temperature phase, in which 
chiral symmetry is broken \cite{Bjo92,Kow92}.  The growth of long 
wavelength modes in the chiral order parameter then occurs quite 
naturally, if the transition proceeds out of equilibrium \cite{Raj93b}.

This picture has been partially confirmed in numerical calculations based
on the linear sigma model \cite{Gav93b,Klu94,Asa94}.  As the coherence 
length is inversely proportional to the growth rate of the instabilities 
\cite{Boy93,Bed93}, larger domains of coherently excited pion field may 
emerge if the chiral order parameter is somewhat, but not far, away 
from its equilibrium value, as is likely to occur in a relativistic heavy 
ion reaction \cite{Gav93c}.  Initial deviations from isospin neutrality 
do not necessarily destroy the usefulness of this probe \cite{Coh94}.  The 
observation of pion charge ratios significantly different 
from ${1\over 3}$, or nonzero charge correlations \cite{Gre93}, would 
therefore be a direct signature of the chiral phase transition.  

Domains of disoriented chiral condensate may also contribute to
antibaryon production through the formation of topological defects in
the chiral order parameter \cite{Deg84,Ell89}.  Such defects can arise
at the intersection of chiral domain walls, which carry 
baryon number and eventually evolve into baryons and antibaryons,
possible leaving a signature of the chiral phase transition in regions
of phase space that are normally baryon-poor \cite{Kap94}.

\subsubsection*{\sc Medium Effects on Hadron Properties}

The widths and positions of the $\rho,\omega$, and $\phi$ peaks in the 
lepton-pair spectrum are sensitive to medium-induced changes of the 
hadronic mass spectrum, especially to the possible drop of vector
meson masses preceding the chiral symmetry restoration transition
\cite{Pis82,Boc84,Dos88,Fur90,Gal91,Aou91,Asa92,Her92,Hat93}.  
In the absence of high baryon density, modifications of the peak positions 
are predicted to be small except in the immediate vicinity of the phase 
transition, whereas the increase in the width of the $\phi$-meson due to 
collision broadening is substantial \cite{Hag93}.  This could serve as a 
measure of the density of the mixed phase \cite{Sei93}.  A change in the 
K-meson mass also would affect the width of the $\phi$ meson 
\cite{Lis91,Bi91}.  A double $\phi$ peak in the lepton pair spectrum would 
be indicative of a long-lived mixed phase \cite{Asa93}.  

\subsection{\it Hard QCD probes}

The color structure of QCD matter can be probed by its effects on the 
propagation of a fast parton \cite{Bjo82,Sve88}.  The mechanisms are similar 
to those responsible for the electromagnetic energy loss of a fast charged 
particle in matter: energy may be lost either by excitation of the 
penetrated medium or by radiation.  

The connection between energy loss of a quark and the color-dielectric 
polarizability of the medium can be established in analogy to the theory 
of electromagnetic energy loss \cite{Tho91,Mro91,Koi91}.  Although 
radiation is a very efficient energy loss mechanism for relativistic 
particles, it is strongly suppressed in a dense medium by the 
Landau-Pomeranchuk effect \cite{Mig57}.  The QCD analog of this effect
has recently been analyzed comprehensively \cite{Gyu94,Pei95}. Adding 
the two contributions, the stopping power of a fully established quark-gluon 
plasma is predicted to be higher than that of hadronic matter.  

A quark or gluon jet propagating through a dense medium will not only lose 
energy but will also be deflected.  This effect destroys the coplanarity
of the two jets from a hard parton-parton scattering with the incident
beam axis \cite{App86,Bla86b}. The angular deflection of the
jets also results in an azimuthal asymmetry.  The presence of a quark-gluon 
plasma is also predicted to enhance the emission of jet pairs with small 
azimuthal opening angles \cite{Pan94}.  The sharp increase in the 
acoplanarity of di-jet events in proton-nucleus interactions recently 
observed at Fermilab \cite{Cor91} indicates that the interpretation of
these signals is complicated by re-interaction.

\section{\rm EXPERIMENT}
\subsection{\it Experimental Conditions}

The techniques used in experiments studying relativistic nucleus-nucleus 
collisions are similar to those used in high energy physics experiments. 
The primary difference is that the particle multiplicities and the 
backgrounds for various processes differ between the nuclear and particle 
physics environments.  For central collisions, with impact parameters near 
zero, the particle multiplicities scale approximately as the mass of the 
colliding system and therefore, with nuclear masses around 200, can be a 
factor of 200 times higher in collisions of heavy nuclei compared to 
collisions between protons at the same energy. The multiplicities scale 
weakly as a function of energy with $dn/dy(y_{\rm cm})\sim \ln(\sqrt{s})$. 
Likewise, the combinatorial backgrounds underlying processes such as
Drell-Yan production, particle and resonance decays, and photon production 
increase more than linearly with (and usually as the 
square of) increasing primary particle multiplicities, 
complicating reconstruction of these signals.

\subsection{\it Present Relativistic Heavy Ion Accelerators}

There are presently two research facilities for relativistic heavy ion 
experiments, focusing on dense hadronic matter
and signatures of quark-gluon plasma formation.
\footnote{Two other relativistic heavy ion facilities at 
somewhat lower energies, which focus on properties of the nuclear equation 
of state, are the GSI-Darmstadt SIS accelerator, and the JINR-Dubna 
Nucleotron.} 
These are the Brookhaven Alternating Gradient Synchrotron (AGS) and 
the CERN Super Proton Synchrotron (SPS), both in operation with heavy ions 
since 1986. Experiments utilize nuclear beams ranging from protons to gold 
with momenta up to $29(Z/A)$ GeV/$c$ from the AGS and from protons to lead 
with momenta up to $400(Z/A)$ GeV/$c$ from the SPS, where $Z$ is the element 
number and $A$ is the atomic mass number of the nuclear beam. These correspond 
to center-of-mass (c.m.) energies per nucleon pair of 4.84 GeV for Au+Au at 
the AGS and 17.2 GeV for Pb+Pb at the SPS. Simply scaling the particle 
multiplicities measured in pp interactions by $A$, the charged particle 
multiplicity density at midrapidity for these systems is approximately 
150 per unit rapidity at the AGS and 270 per unit rapidity at the SPS. 

In these experiments the collisions occur with a target in the laboratory 
frame, in contrast to colliding beams. The products are focused toward 
forward angles in the laboratory. For example, mid-rapidity ($\theta_{\rm cm} 
= 90^\circ$) corresponds to $\theta _{\rm lab} \leq 20^\circ$  and 
$\theta_{\rm lab} \leq 5.5^\circ$, at the AGS and SPS respectively, for the 
majority of particles emitted with less than 1 GeV/$c$ transverse momentum. To 
study particles emitted in the mid-rapidity region, the experiments require 
compact, highly segmented detectors placed in the forward regions downstream 
from the target. The particle momenta are relatively high due to the Lorentz 
boost of the center-of-mass into the laboratory frame 
making particle identification 
via ionization energy loss and time-of-flight more difficult. Particles at 
mid-rapidity have momenta in the relativistic rise region, thus energy loss 
measurements require larger numbers of samples along tracks \cite{Wal79}
and exceptional track separation capabilities.

\subsection{\it Heavy Ion Colliders}

There are two colliders planned for acceleration of heavy ions to 
ultra-relativistic energies. The Relativistic Heavy Ion Collider 
(RHIC) \cite{RHI89}, 
presently under construction at Brookhaven National 
Laboratory in New York, is a dedicated heavy ion collider 
planned for experiments in 1999. RHIC will accelerate and collide 
ions from protons to heavy nuclei, such as Au at c.m. 
energies up to 500 GeV for protons and 200 GeV per nucleon-pair for Au 
nuclei. The luminosity for Au+Au will be 2 x 10$^{26}$ cm$^{-2}$ s$^{-1}$. 
Near head-on collisions of Au+Au at RHIC are expected to produce from 500 to 
1500 charged particles per unit pseudorapidity at midrapidity in a single 
collision. Large detector systems are being constructed \cite{RHI93} to 
analyze the products of these interactions for formation of a quark-gluon 
plasma and a possible chiral phase transition.

Heavy ion physics research \cite{Sch94} will also be an integral part of the 
program for the Large Hadron Collider (LHC), to be constructed at CERN, 
the European Centre for Nuclear Physics, in Geneva, Switzerland.  For Pb 
nuclei, the c.m. energies at the LHC will be 5.4 TeV per nucleon pair with 
luminosities of 10$^{27}$ cm$^{-2}$s$^{-1}$. Predictions for the charged 
particle densities at the LHC for near head-on collisions of Pb+Pb range 
from 2000 to 8000 per unit pseudorapidity. The large uncertainty in these 
numbers arises primarily from the present lack of information on the
distributions of soft gluons in nuclei \cite{Hera95}.

\subsection{\it Detector Components}

The types of detectors anticipated for use in the collider experiments can 
be divided into four categories: detectors for charged particle tracking,
calorimeters for energy measurements; detectors for particle identification;
and photon detectors.  In contrast to the high energy physics environment,
at the heavy ion colliders 1) the $p_T$ 
of the particles of interest is typically lower; 2)
the luminosities are considerably lower,
allowing the use of slower detectors and readout times; while 3) the 
particle multiplicities are considerably higher requiring finer segmentation 
of detectors and larger event sizes.

Tracking detectors utilize the 
ionization of a charged particle traversing a medium in order to determine 
its trajectory.  For tracking near the primary collision region within 5 to 
10 cm, where particle densities approach $\sim$ 100 to 1000 cm$^{-2}$,
silicon detectors (pixels, strips, drift) \cite{Lut95} with excellent 
position (20$\mu$) and double track (200$\mu$) resolution are used.
Measurements close to the primary interaction are particularly important 
for detecting decays of short-lived strange and charm particles, of extreme 
importance in quark-gluon plasma searches.  For large area tracking away 
from the interaction region and at more moderate particle densities of 
$\sim$ 1 cm$^{-2}$, time-projection chambers and other types 
of tracking detectors are used \cite{Blu94}.

The calorimeters used at the heavy ion colliders will be of two basic types.
Conventional sampling calorimeters \cite{Fab87} can be used for 
electromagnetic and hadronic energy determination, and measurements of jets.
Highly segmented calorimeters can be used, in addition to the above 
measurements, to measure high energy particles and photons. New types of 
calorimeters \cite{Ell94} with fine segmentation and various types of 
readout have recently been designed and tested for use in the high track 
density environments at heavy ion colliders.

Particle identification of charged particles can be accomplished using
ionization energy loss, Cerenkov radiation, transition radiation, or
time-of-flight techniques. At higher momenta, combinations of these
techniques are sometimes necessary for best results, especially when
measuring over a wide range of $p_T$ over which any single technique
may not be applicable. 

Highly segmented photon detectors will be utilized for the measurement of 
photon radiation. Detectors from new types of materials have been developed 
\cite{Kie91} for higher efficiencies and with smaller 
Moli\`ere radius to be able to improve performance and to more finely 
segment photon detector systems.

\subsection{\it Detector Systems and Measurements}

There are various types of detector systems that will be utilized at 
the heavy ion colliders. For illustration, we will briefly point out three,
and describe the measurements they allow.

\subsubsection*{\sc Hadron Spectrometers}
Hadron spectrometers utilize tracking detectors, detectors for particle 
identification and magnetic fields to determine particle types and their 
momenta. Such spectrometers measure momentum (and rapidity) distributions 
for a large variety of identified particles, including decays of particles 
and resonances.  For larger acceptances particle correlations can be measured. 
The PHOBOS \cite{PHO94} and BRAHMS \cite{BRA95} experiments planned for RHIC 
are examples of these.  For very large acceptances, approaching complete 
solid angle coverage, measurements of single event observables can be 
performed. The STAR experiment \cite{STAR92} under construction for RHIC and 
the ALICE experiment \cite{ALI94} planned for the LHC are examples.  Such 
measurements will be unique to the high multiplicity heavy ion experiments. 
Examples of single event observables are the strangeness content; temperatures 
of pions and kaons derived from the spectra or mean transverse momenta; event 
shapes and source sizes; and energy, momentum and particle number fluctuations 
as a function of emission direction.  The purpose is to link these observables
to thermodynamic variables and other dynamical properties of the evolving 
system and thus gain information on the dynamical evolution and state of the 
system.  An additional aspect of large acceptance spectrometers is the 
ability to measure jets.

\subsubsection*{\sc Lepton Pair Spectrometers}
Lepton pair spectrometers typically measure $e^+e^-$, $\mu^+\mu^-$, 
and $e\mu$.  They focus primarily on the lepton-pair mass spectrum over the 
entire mass range available, leptonic decays of hadronic resonances 
(including the $J/\psi$ and higher mass resonances when possible), and the 
Drell-Yan background.  Here again a magnetic field is utilized for momentum 
measurements along with arrays of detectors for tracking and particle 
identification. The PHENIX experiment \cite{PHE93} under construction for 
RHIC is an example.  
Besides using various techniques to distinguish
particle pairs from the decays
in the primary heavy ion interaction, it is also 
important to differentiate
between particle pairs produced in the primary 
interaction and those produced in secondary processes away from the primary 
vertex. Since the yields of higher mass pairs are low and the 
branching ratios for the electromagnetic decays to $e^+e^-$, $\mu^+\mu^-$, 
and $e\mu$ are small $(1/137)^2$ compared to hadronic decays modes, 
suppression of hadronic background sources relative to the electromagnetic 
signals (by factors of $\leq 10^{-4}$) requires excellent tracking and 
particle identification using various combinations of detectors.

\subsubsection*{\sc Photon Spectrometers}
Highly segmented photon spectrometers seek to measure direct photon 
radiation from the the various phases of matter formed in heavy ion 
collisions.  Such measurements require very elaborate detection
systems to suppress the photons from neutral pion decay and electron
conversion in order to measure the direct photons.  The extraction of 
weak direct photon signals from the background requires a detailed 
understanding and careful analysis of systematic errors, understanding 
and subtraction of the combinatorial background, and an understanding 
of the sensitivity to the various decay backgrounds. In addition to the 
efficiencies, the multiplicity dependence of the identification of the 
various particles creating the backgrounds must be understood. 

\section{\rm PRESENT UNDERSTANDING OF DATA}

The quark-gluon plasma has yet to be uniquely observed or identified. 
There are, however, experimental observations that require for their 
description more than our present understanding of the standard model of 
hadronic interactions. Some require modifications to existing microscopic 
models, due primarily to the high density environment of these reactions, 
others are more easily described by models incorporating deconfined quarks 
and gluons. In this section, we will place into perspective the present 
understanding of the data and review briefly new features which affect 
this understanding. An emphasis on the QGP signatures and those observations 
which require more than hadronic interactions for their description will be 
made.

The status of the experimental results at these center-of-mass energies,
$\sqrt{s} = 5-20$ GeV, was last summarized in this Journal in 1992
\cite{Sta92}. Since that time there has been significant progress in
understanding the interactions of nuclei at high energy. This is a result 
of recent sophisticated measurements, availability of heavy nuclear beams,
accumulation of systematic data, and improved theoretical model calculations.

\subsection{\it Baryon and Energy Stopping}

The nuclear stopping power is a measure of the degree to which the 
energy of relative motion of two incident nuclei can be transferred into 
other degrees of freedom. The amount of nuclear stopping determines basic 
parameters, such as energy and volume of the interaction region and thus 
energy density, which govern the reaction dynamics and the extent to 
which conditions are favorable for the formation of a deconfined phase. 
Experiments determine the stopping power of colliding nuclei from
1) the redistribution of the incident protons
into proton final-state rapidity distributions;
2) measurements of the energy remaining in the forward-going baryons,
which carry the initial energy into the reaction;
and 3) measurements of the transverse energy distributions 
which represent the energy transformed into produced particles and their 
kinetic motion.

Many systems have been studied at the AGS and the SPS to determine the 
nuclear stopping power.  
The rapidity distributions for protons from pp interactions \cite{Blo74}
and peripheral nucleus-nucleus interactions 
at the AGS \cite{Abb94,Bar94a} and SPS \cite{Bae94}
energies are peaked forward and backward in the c.m. frame, near the 
projectile and target rapidities, exhibiting a small degree of stopping.
For central collisions of two intermediate mass nuclei ($A \sim 30$)
the proton rapidity distributions spread over the entire rapidity 
space with broad peaks approximately half-way between the target/projectile
and the c.m. rapidities, thereby 
exhibiting a fairly large amount of stopping at both energies
\cite{Abb94,Bar94a,Bae94}.  The distributions for the 
heavy systems (mass $A \sim 200$) at both energies peak 
and exhibit a ``pile-up'' of matter at midrapidity \cite{Vid95,Dod95}.
The heavier systems are 
more efficient at stopping the incoming matter and thus higher energy 
densities are expected to be reached when colliding these heavier nuclei. 

The measured rapidity distributions of protons and produced particles at 
the AGS energy can be reproduced by the ARC cascade model \cite{Pan92}. 
This model describes the nuclear reaction as a sequential binary cascade
of interactions among known hadrons, neglecting all medium effects on the 
hadron-hadron cross sections. This is not sufficient at the higher SPS 
energies, especially for the proton distributions.  The rapidity 
distributions of protons from central collisions of 160 A-GeV/$c$ Pb+Pb 
exhibit a considerably higher degree of stopping than predicted by models 
based on binary hadron interactions with free-space cross sections, such 
as HIJING \cite{Wan91}, VENUS \cite{Wer93a}
and FRITIOF \cite{And87}.  There apparently exist 
important mechanisms for the transfer of baryon number that are not 
incorporated in binary cascade models.  In order to successfully describe 
the measured rapidity distributions of baryons at SPS energy, cascade 
models have had to incorporate novel reaction mechanisms involving 
multi-hadronic intermediate states, such as color ropes in RQMD \cite{Sor92}
or multi-quark clusters in VENUS \cite{Aic93}.  The presence of these 
effects must be viewed as a result of the high density.  They show that 
nuclear collisions at SPS energies require a description that goes beyond 
conventional hadronic interaction physics. 

The transverse energy E$_T$ distributions have also been measured to determine the
degree of nuclear stopping in central heavy ion collisions at the AGS and SPS. 
Measurements of E$_T$ at the AGS \cite{Bar93b} have shown that the E$_T$ 
increases fifty percent more rapidly, in going from Si + Al to Au + Au,
than predicted from an independent nucleon-nucleon
interaction model. An interpretation of this increase in a microscopic model
(ARC) \cite{Pan92} is that there is a considerable increase in the baryon 
density, up to ten times normal nuclear density, and
an accompanying large increase in the volume of high density
matter when going from Si+Al to Au+Au at the AGS \cite{Bar94b}.
At the SPS, measurements of E$_T$ in central collisions
of 160 A-GeV/$c$ Pb+Pb \cite{Alb95a}
exhibit a large amount of energy transfer into particle production.
The accompanying energy density was estimated to be 3 GeV/fm$^3$,
similar to the results obtained for 200 A-GeV/$c$ S+Au central collisions 
\cite{Bae91} but over a larger volume. 

The overall degree
of nuclear stopping is found to be large
in collisions of heavy systems
at both the AGS and SPS energies. There is a significant
build-up of baryon and energy density over a large volume
and considerable energy transfer
from the initial relative motion into
particle production, suggesting that conditions 
may be favorable for thermal and chemical equilibrium.

\subsection{\it Thermal Equilibration -- Temperatures and Flow}

The application of thermodynamic concepts to multiparticle production has 
a long history.  Strictly speaking, the concept of temperature applies only 
to systems in thermal equilibrium with a heat bath. In high energy 
interactions the kinetic energy of the longitudinal motion, along the beam 
axis, serves as an energy reservoir. Thermalization is normally only thought 
to occur in the transverse degrees of freedom.  The transverse momentum 
distributions measured in nucleus-nucleus collisions reflect conditions at 
the time when interactions cease (freeze-out).  Due to the large 
re-interaction cross sections for hadrons the distributions do not reflect 
earlier conditions, as in a hot and dense deconfined phase, when chemical 
and thermal equilibrium may have been established.  It is therefore not 
feasible to extract a temperature from the slopes of hadron spectra 
and relate to it an observable of the high density phase, such as entropy 
or energy density. Prior to this a better understanding of the effects on the 
spectra of final state interactions and collective flow is necessary.

A discussion of the systematics of the slopes of the spectra of various 
particles and colliding systems measured at the AGS and SPS has been 
presented elsewhere \cite{Sta92}.  The momentum spectra for the light, 
produced particles are not thermal, and have significant deviations both 
at low $p_T$ ($\leq$ 0.25 GeV/$c$) and at high $p_T$ ($\geq$ 1.0 GeV/$c$). 
A collective nuclear flow component\footnote{Originally observed 
in relativistic nuclear interactions at the Bevalac \cite{Bevflow}.}
has been measured in azimuthal energy distributions at the AGS \cite{Bar94c}.
This evidence for collective 
flow clearly underlines the need for a better understanding and description 
of the phenomenon. It is also necessary to incorporate collective flow 
in models and to be able to subtract its effects in any extraction of 
temperatures from spectra.  The measured transverse momentum and transverse 
mass distributions are consistent with a thermal description at freeze-out, 
only if additional collective radial flow and feeding from 
higher-lying resonances are taken into consideration.\footnote{Note 
that these effects cannot be explicitly separated in the spectra alone.}

Data measured at the AGS for 14.6 A-GeV/$c$ Si+Au \cite{Abb94} are 
compatible with thermal distributions including transverse flow with 
temperatures $120<T<140$ MeV and mean transverse flow velocities 
0.33 $\leq\langle\beta\rangle\leq$ 0.39 over this range of temperatures
\cite{Bra95a,Let94b}.  Thermal fits including resonance decays and transverse 
flow for SPS data in central S+W collisions are consistent with 
T = 160 MeV and a transverse flow with average velocity $\langle\beta\rangle
\approx$ 0.41 \cite{Bra95b}. The central S+S data are compatible with 
T = 150 MeV and an average velocity $\langle \beta \rangle \approx$ 0.32
\cite{Sol91}. These results are consistent with the emitting
system being in thermal equilibrium at freeze-out.\footnote{The 
fits presently do not rule out a much higher temperature ($T=190-230$ MeV)
combined with the absence of transverse flow, in the case of the SPS data
\cite{Raf91,Sol94}.  A thermal freeze-out at apparent temperatures as high 
as T = 230 MeV is inconsistent with a {\it hadronic} picture at freeze-out
if the results from the lattice gauge theory on T$_c$ are correct. In this 
case hadron emission would have to proceed far off equilibrium, 
possibly reflecting thermal conditions in a pre-hadronic phase of the reaction.}

\subsection{\it Chemical Equilibration -- Strangeness Production}

The consistency of the assumption of particle emission from a locally 
thermalized source can be tested by using a thermo-chemical model to 
describe the ratios of the various emitted particles.  This yields a 
baryon chemical potential $\mu_B$, a strangeness chemical potential $\mu_s$ 
and a temperature $T$ at chemical freeze-out,
which can be compared to the temperatures derived from the particle spectra.
Of specific interest are strange particles and anti-baryons, whose 
production is predicted to be enhanced \cite{Raf82a,Koc86} if a
chiral phase transition occurs in a dense, baryon-rich system. 

An enhancement in the production of strange particles has been observed 
and measured for various systems in nucleus-nucleus collisions at the 
AGS \cite{Abb94} and the SPS \cite{Alb94a,And92,Aba94,Kin95,Gaz95}, compared to 
proton-proton and proton-nucleus interactions. It should be noted that some 
increase in the production of strange hadrons, especially kaons and 
$\Lambda$-hyperons, in nucleus-nucleus collisions can result from purely 
hadronic interactions \cite{Mat86b,Lee88,Mat89}.  
However, the enhancement in the 
$\Lambda$ yield measured over a large rapidity interval \cite{Alb94a} is 
difficult to describe by a cascade of hadronic interactions.  A clear 
enhancement in the production of $\overline{\Lambda}$, $\overline{\Xi}$, 
$\Omega$ and $\overline{\Omega}$  hyperons \cite{And92,Aba94,Kin95} has been 
measured at the SPS.  To be able to describe this enhancement severe 
modifications must be introduced to the hadronic cascade models invoking 
precursors of quark-gluon plasma formation such as creation of color 
ropes \cite{Sor92}, breaking of multiple-strings \cite{Wer93b}, or decay of 
multi-quark droplets \cite{Aic93}. 

The ratios $K^{+}/K^{-}$, $\overline{\Lambda}/\Lambda$ and $\overline{p}/p$ 
from the AGS are consistent with an equilibrated hadronic fireball with 
$\mu_{s}/T = 0.54\pm 0.11$ and $\mu_{B}/T = 3.9\pm 0.3$
\cite{Let94b,Raf94} (see also \cite{Bra95a}). 
The $\overline{\Lambda}/\Lambda$, $\overline{\Xi}/\Xi$, $\Xi/\Lambda$ and
$\overline{\Xi}/\overline{\Lambda}$ ratios \cite{Aba94} measured at the SPS 
have been reproduced in an equilibrium hadron gas description 
with $\mu_{s}/T = 0.24-0.28$ and $\mu_{B}/T = 1.05$ \cite{Bra95b}.
The SPS strange particle data have also been described by an equilibrated 
quark-gluon plasma, with strangeness neutrality ($\mu_{s}$ = 0) and 
large strangeness saturation ($\gamma_{s} \geq 0.5$)
which hadronizes and decays instantaneously \cite{Let92,Let93a,Let93b}.
Further experimental information is necessary to differentiate
between the hadron gas and quark-gluon plasma descriptions of the
strange particle ratios.

\subsection{\it Two Particle Correlations}

Two-particle interferometric techniques \cite{Zaj93} have been used to study 
the space-time evolution of the colliding system and to provide information 
for testing and understanding dynamical models. Pions and kaons will freeze 
out late in the expansion and will provide information on the system in its 
later stages.  In experiments at SPS energies the radii extracted from 
$K^{+}K^{+}$ and $K^{-}K^{-}$  interferometry measurements \cite{Bek94} are 
the same indicating that the interactions of the kaons with the expanding 
matter are predominantly $K\pi$ interactions \cite{Mur94}.  A similar 
comparison has yet to be made at the AGS, where the central rapidity
region has approximately equal numbers of mesons and baryons.  At AGS and 
SPS energies, the radii derived from $\pi\pi$ interferometry are 
consistently larger than those from KK measurements \cite{Bek94,Sul93}.
This has been attributed \cite{Sul93} at the higher energies to
effects of resonance decays  \cite{Gyu88} in addition to the differences in
$\pi$-hadron and K-hadron interaction cross sections. 

Studying the source sizes derived from correlation measurements of each 
of the particle species provides information on the space-time evolution 
of the expanding source. 
The longitudinal and transverse radii measured 
in central collisions at the SPS are considerably larger than the 
projectile radius, thus reflecting a large amount of expansion of the 
system in the final stage before freeze-out \cite{Jac95a}. A similar, but 
less pronounced, effect is observed in experiments at the AGS. The 
longitudinal source radii measured as a function of rapidity \cite{Alb95b}
can be fit by a boost invariant longitudinal expansion \cite{Mak88}.

A slight decrease in the transverse source radii as a function of 
increasing transverse momentum has been measured in sulphur-induced 
collisions at the SPS \cite{Alb95b}.  A considerably larger decrease has
been observed for the heavy Pb+Pb system at the SPS \cite{Alb95c}.  The 
transverse source radii are expected to decrease as a function of increasing 
transverse momentum in the presence of transverse flow \cite{Pra86}.

In a detailed comparison of a model calculation with the measured data, 
a correlation is observed between a particle's momentum and position at the 
time of freeze-out \cite{Jac95a}.  This suggests that the assumption of the 
absence of kinematic correlations required for the correlation function 
analysis is not completely realized and that shadowing and 
flow effects are important to include in interpreting the data. 

An overall picture of the source in the late stage of the collision 
emerges from the $\pi\pi$ correlation measurements. It is one of a 
pion-emitting source which expands longitudinally in a boost invariant 
way, with a slight transverse expansion which increases with the mass of 
the colliding system. The volume of the expanding source at freeze-out is 
proportional to the rapidity density of produced particles, suggesting 
that freeze-out occurs at a constant density \cite{Sta92,Gaz95}.  At SPS 
energies the time between the onset of expansion and freeze-out along the 
longitudinal direction is approximately 5 fm/$c$ and a  small 
difference between the transverse components is observed indicating a short 
duration, not more than 2 fm/$c$, for particle emission \cite{Fer92}.  
To date, all 
measurements of the two transverse radius components of the source, which are 
distinguishable if a long-lived intermediate phase were to exist, have been
identical. Thus no long-lived intermediate state has yet been observed in the
correlation measurements \cite{HBT95}.

\subsection{\it Resonance Matter}

At temperatures and densities just below that of a quark-gluon plasma,
nuclear matter is expected to exist in the form of highly excited resonance 
matter.  Microscopic calculations \cite{Pan92,So92b} predict that during 
central collisions at AGS energies, the central rapidity region becomes 
compressed to high densities, is highly excited and baryon-rich
\cite{Sto86,Sor90,Sch92}.  A significant fraction of the baryons in central 
collisions at AGS energies will be in excited states 
\cite{Pan92,Sor91,Hof94}, forming what might be 
called baryonic resonance matter.  This is 
consistent with a recent measurement of the
$\Delta$ resonance population at the time of freeze-out 
in central collisions at the AGS \cite{Hem94,Bar95}
and results of RQMD, which find that
35 percent of the final-state nucleons are excited
to the $\Delta$ resonance at the time of freeze-out.
Excitation to higher energies at the SPS, RHIC and the LHC
should lead to the formation of highly excited
resonance matter. Excitation of higher-lying resonances provides a 
means of converting energy from the relative motion thereby
increasing the rate of equilibration. Studies of the population of 
the various resonances in these collisions will
provide a test of thermalization and should
distinguish features of the collision dynamics 
between the AGS and SPS energies.

\subsection{\it Virtual and Real Photons}

An enhancement has been measured in the invariant mass spectrum of muon 
pairs emitted in central nucleus-nucleus collisions relative to proton-proton 
and proton-nucleus interactions at 200 A-GeV/$c$.  The observed yields in 
nucleus-nucleus collisions exceed the contributions from known sources
(combinatorial background, Drell-Yan, open charm and 
hadronic decays) \cite{Lou94,Mas95}
over the range of invariant masses $0.2 < M < 2.5$ GeV/$c^2$ (up to the 
onset of the $J/\psi$ for which a suppression is observed) and for all 
transverse masses, whereas the proton-nucleus data are well
described by the same known sources.  A large 
excess is also measured in the low-mass region $(0.2 < M < 1.5$ 
GeV/$c^2$) for $e^{+}e^{-}$ pairs in S+Au collisions relative to p+
Be and p+Au interactions at 200 A-GeV/$c$ \cite{Aga95}.  The spectra for the 
proton-induced reactions are well-reproduced by $e^{+}e^{-}$  pairs from 
known hadronic sources.  The enhanced spectra can be described assuming 
medium modifications to the intermediate mass vector meson resonances due 
to partial chiral symmetry restoration \cite{Li96}.

Direct photon measurements have been made using nucleus-nucleus collisions
in three separate experiments at the SPS and preliminary results have
been reported. No direct photons were observed at the level of approximately
10\%\ systematic error in one experiment \cite{Aga95} and slightly larger
systematic error in another \cite{Ake90}. A third experiment \cite{Awe95} 
has reported a 5.8\%\ photon signal over background with a 5\%\ systematic
error. Thus preliminary experimental results are consistent with one another 
within experimental errors, and a significant direct photon yield has yet 
to be established. 

\subsection{\it $J/\psi$ Suppression}

A suppression of $J/\psi$ \cite{Bag89} and $\psi'$ \cite{Abr94} production 
relative to that of the Drell-Yan continuum has been measured for central 
collisions in nucleus-nucleus experiments at the SPS. The suppression of the 
$\psi'$ is observed to be larger than that of the $J/\psi$ in central 
nucleus-nucleus collisions. Such suppression \cite{Mat86} is predicted to 
result from color screening of the $c\overline{c}$ pair in a deconfined 
medium. It has also been predicted to occur as a result of final-state 
interactions in a dense hadronic medium \cite{Gav94}.  A similar suppression 
has also been seen in $J/\psi$ production in hadron-nucleus interactions 
\cite{Ald91} and $\mu$-nucleus interactions \cite{Ama91}, lending credence 
to a hadronic mechanism to describe the observed suppression.  However, 
$\psi'$/$\psi$ ratios have been measured in proton-proton and proton-nucleus 
interactions and found to be constant, independent of the nuclear mass of 
the target \cite{Ram95}.

There is presently 
no unambiguous explanation for the suppression of $J/\psi$ and $\psi'$ in 
the various interactions in which they have been measured. However, 
theories describing the suppression of $J/\psi$ and $\psi'$ in 
nucleus-nucleus collisions require the formation of comoving 
high-density matter, be it hadronic or deconfined matter.  
Better knowledge of the various mechanisms for interactions
of the $J/\psi$ in hadronic matter and deconfined matter, as well as initial
and final state interactions, is important to 
understanding the suppression.

\section{\rm FUTURE MEASUREMENTS}

\subsection{\it Kinematic Probes}

Future experiments will study properties of high {\it baryon} density in 
collisions of the heaviest systems at the AGS and SPS, and high {\it energy}
density with heavy systems at RHIC and LHC.  More information on collective 
flow at the AGS and SPS will emerge from the comparison of correlation 
measurements with single particle spectra and from calculations using 
microscopic models.  The goal of these investigations will be to better 
understand the effects of flow on other observables. This will be essential 
for a knowledge of the space-time evolution of the system and subsequently
for determining the entropy, temperature and chemical potentials of the system
in the high density phase. Furthermore, the energy dependence of strangeness 
and entropy production should be measured in the energy regime from the AGS 
through the SPS energies, to understand the presently observed differences in 
the strangeness and entropy at the two energies and to search for a possible 
onset of a phase transition at moderate to high baryon density. 

Studies with the heavy systems, especially at higher energies, will provide 
unprecedented energy densities, where effects of the quark-gluon plasma and 
chiral transitions are expected.  One of the most interesting characteristics 
of the new experiments will be to make use of the increased multiplicities
in the heavy systems at SPS and higher energies to extract thermodynamic 
properties (T, $\mu_B$, S, $\mu_s$, $\epsilon$, etc.) of the system on an 
event-by-event basis \cite{NA49,STAR92}.  This enables categorization of 
individual events into groups according to thermodynamic properties and could 
potentially lead to the isolation of events with special properties associated 
with quark-gluon plasma formation.  Ensembles of these events could then be 
studied in greater detail to determine their particular characteristics.

\subsection{\it Electromagnetic Probes}

New measurements using electromagnetic probes to study heavy colliding 
systems at higher energy densities, at the SPS and at the higher energies 
of RHIC and the LHC, should provide important new results.

The observation of an excess of lepton pairs in the low and intermediate 
mass region from three different experiments at the SPS requires further 
experimental and theoretical investigation.  There are differences in the 
measurements which must be understood.  Better statistics are needed to be 
able to understand the dependence of the excess on invariant mass and on 
collision centrality.  For low mass electron pairs, there appears to be
a threshold at $M = 2m_{\pi}$, a quadratic dependence on rapidity density,
and the excess is observed over a broad mass range. These facts suggest that 
the excess may be due to $\pi^{+}\pi^{-}$ pair annihilation in a dense medium.
The excess above the $\rho$, which is similar to the open charm contribution, 
might be explained by enhanced production of charm.  New results using the 
heavy systems at the SPS should provide an even stronger enhancement of the 
lepton-pair yields, if the excess seen so far is due to effects of a dense 
hadronic medium.  At RHIC and the LHC, higher mass lepton pairs with masses 
between those of the $J/\psi$ and $\Upsilon$ could be used as a diagnostic 
tool for the collision dynamics \cite{Kap92}.
Measurements of the ratios of the $\rho$, $\omega$, and $\phi$
vector mesons will provide information on the time evolution of 
collisions and, in particular, the lifetime of the fireball.

In measurements of direct photons, better control of experimental
conditions is necessary in order to reduce the systematics
to the level required for measuring signals only a few percent
above background. Theoretically, it will be very important to 
understand the relationship between the large excess of low mass electron 
pairs (virtual photons) and a weak direct photon signal.

Since the transverse mass and momentum distributions of hadrons contain 
effects due to resonance production, decays, expansion, and collective 
flow, it would be extremely interesting to utilize noninteracting 
probes, such as photons, to measure the temperature of the high density
phase. Clearly, a consistent picture between these results and the 
temperature derived from the observed 
particle ratios would be both gratifying and most convincing.

\subsection{\it Probes of Deconfinement}

Further studies of the relative suppression of $J/\psi$ and $\psi'$ 
at the SPS in central collisions of very heavy nuclei
should provide crucial information on the relative roles of nuclear 
rescattering, color screening and deconfinement
at the highest energy densities.  Measurement of $J/\psi$ and $\psi'$ 
in A+p interactions using reverse kinematics to measure the $J/\psi$-hadron
rescattering cross sections will be important to understanding
the role of rescattering in the break-up of the $J/\psi$ and $\psi'$. 
This is an essential step toward isolating the suppression 
due to screening in a quark-gluon plasma.  It will be important to vary 
the projectile and target masses to determine the A-dependence of 
the various processes for a complete understanding of the
suppression.

At the higher incident energies of RHIC and the LHC, 
the energy densities will be 
more than a factor of ten greater than in
present measurements at the SPS. Thus
the differences in the $J/\psi$, $\psi'$,
$\Upsilon$ and $\Upsilon'$ suppression 
and the screening due to quark-gluon
plasma formation should be accentuated.

\subsection{\it Probes of Chiral Symmetry Restoration}

Effects of a high density nuclear, hadronic or deconfined
medium on the mass and width of the light vector mesons
will be investigated
in central collisions of
very heavy ions at the AGS, SPS, RHIC and the LHC.
This should provide information on
possible resonance mass and width modifications due to 
the high density medium and chiral symmetry 
restoration. Another characteristic of 
a chiral phase transition would be the observation of 
abnormal ratios of charged to neutral pions. This will
be investigated in event-by-event measurements
at higher densities in central collisions of the
heavy systems at the SPS, RHIC and LHC.

\subsection{\it Other Probes}

Because a significant fraction of the interactions at RHIC and the
LHC will be in the perturbative regime, it
is important to measure the distributions of the incident partons 
in nuclei and the nuclear shadowing in collisions of nuclei.
Deep inelastic scattering of leptons from nuclei has provided measurements of
the quark distributions in nuclei.  However, direct access to the gluon 
distributions will require measurements of p+A interactions. 
Comparisons of similar measurements in A+A interactions
will provide the understanding of nuclear shadowing
necessary to calculate these interactions at the microscopic level.

It would be extremely interesting to measure open charm at the SPS and 
at higher energies. Since the initial stages of the collisions
at RHIC and LHC are expected to be dominated by gluons, 
open charm production is expected to be enhanced. Measurement of 
the amount of open charm will provide information on the initial stages
of the collision, prior to formation of a quark-gluon plasma.

Collisions at RHIC and the LHC will exhibit effects of hard scattering of 
partons. Such QCD hard scattering
processes will result in the production of high $p_T$ particles 
and jets which will be measured in experiments
at RHIC and the LHC to test the 
propagation of high $p_T$ partons in highly excited matter and
a quark-gluon plasma. 

New and more sensitive searches for the H-particle 
(neutral, doubly-strange dibaryon) and
strangelets have begun at the AGS \cite{Kum95} 
and SPS \cite{Dit95,Bor94}. These experiments
seek to provide the first spectacular observations of such particles
or at a minimum to set stringent limits on their production cross sections. 
Such measurements would provide important input into model calculations
for the quark structure of 
nuclei, quark-gluon plasma formation, cosmology and stellar evolution.

\section{\rm SUMMARY}

In this review we have described the various signatures 
that have been proposed for quark-gluon plasma formation and chiral
symmetry restoration. These signatures are being pursued vigorously
in experiments at the present day relativistic heavy ion accelerators,
the AGS and SPS, and in the construction of relativistic heavy ion 
experiments for RHIC and the LHC. 
Figure 3 provides an overview of the various signatures
that have been described in this review and their expected behavior
as the energy density (as measured by the transverse energy density)
increases through the critical energy density $\epsilon_c$.
Large changes are anticipated
as the energy density of the transition is traversed.

The searches for signatures of the quark-gluon plasma at the AGS and SPS
have provided an interesting 
initial study of the behavior of these signatures
as the critical energy density is approached from below.
To date, no unambiguous signal of the quark-gluon plasma
or the chiral transition has been seen.
However, several experimental results 
are incompatible with predictions of models
based on established knowledge about interacting hadrons. 
In particular, we note the observation of 
a substantial enhancement of strange (anti-)baryon yields and an
increased emission of lepton pairs in the mass region below the $\rho$-meson.
These results may indicate that a quark-gluon plasma or mixed phase is
formed in some interactions or regions of interactions
at SPS energies, or that hadron masses are substantially reduced
during these reactions. It will be up to future investigation to 
determine this.

Analysis of the observed hadron spectra and yields has provided 
evidence that the system, at the moment of break-up, is in
a state of local equilibrium not far from the predicted phase boundary
between hadron and quark matter, corresponding to $T\approx 160$ MeV,
$\mu_B \approx 170$ MeV at the SPS and $T\approx 130$ MeV, $\mu_B \approx 
500$ MeV at the AGS. If the present theory and models
are correct, and equilibrium is reached at an even earlier stage 
(than freeze-out) of the reaction, 
at a minimum a quark-hadron mixed phase is 
formed at the SPS. This spells exceptional promise
for experiments at the future heavy ion colliders, where the
full specter of quark-gluon plasma signatures can be measured, 
many on an event-by-event basis.

The success of thermal models in describing the rapidity and transverse 
momentum spectra, and ratios of various particles has led to a general 
understanding of the reaction dynamics at these energies. The freeze-out 
temperatures that have been measured are in the range $T = 120-160$ MeV, 
using best fits to the transverse momentum spectra which require a moderate 
amount of transverse flow, $\beta \approx 0.4$.  The quark chemical 
potentials are in the range from $\mu_{q} = 50-70$ MeV for the SPS 
energies and $\mu_{q} = 150-200$ MeV for the Brookhaven AGS energies. 
These measurements can be used to determine the region of the phase diagram 
of nuclear matter that is being investigated in these interactions. The 
resulting data are shown superimposed on a schematic phase diagram of 
nuclear matter in Figure 4. In studying this diagram, it is important to
keep in mind that the freeze-out temperatures are those observed in the
particle spectra after interactions cease. In deriving these temperatures
effects of flow and resonance production and decay have been taken into
consideration. Temperatures at earlier times during the interaction must
have been higher. This has been seen in the results of microscopic,
hadronic cascade calculations. 
The collision evolves from initial nuclear 
densities and temperatures to higher densities and temperatures prior to 
expansion and cooling (denoted by the arrows for the
AGS and SPS measurements) to the final freeze-out values observed in the
experiments. 

It is the goal of the field of heavy ion physics to 
explore the various regions of the phase diagram of nuclear matter (Figure 4)
and to map out its properties at these various temperatures and pressures.
It is of special interest to relativistic heavy ion physics
to investigate regions of higher temperatures and densities
for formation of a quark-gluon plasma and a chiral phase transition.
The new relativistic heavy ion collider experiments
will search for the signatures (Figure 3) of these phase transitions,
and measure observables which reflect variables 
of the state of the system, in order to 
determine the characteristics of nuclear matter
at high densities.

\section{\rm ACKNOWLEDGEMENTS}

We thank T. Hallman, G. Young, and W. Zajc for comments on the text.
We are indebted to S. Nagamiya for the original concept of Figure 3.
JWH is grateful for the support of the Alexander von Humboldt Foundation 
and the hospitality of the Institut f\"ur Kernphysik of the Universit\"at 
Frankfurt. This work was supported in part by the 
Director, Office of Energy Research, Division of Nuclear Physics of the Office 
of High Energy and Nuclear Physics of the U.S. Department of
Energy under Contract DE-AC03-76SF00098 and Grant DE-FG05-96ER40945.

\newpage

\begin{figure}
\centerline{\psfig{figure=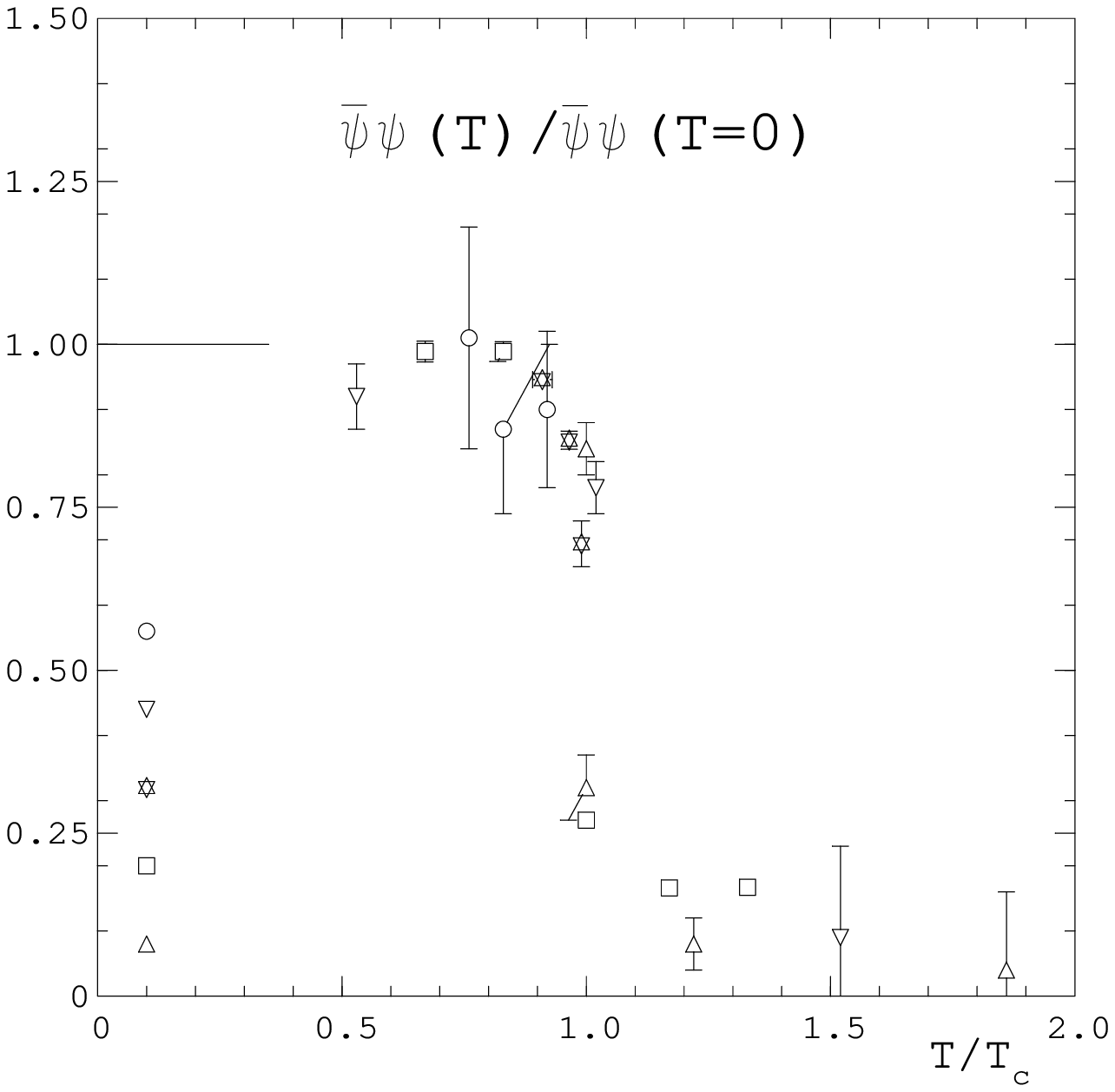,width=5in,angle=-90}}
{\small Figure 1: Quark condensate $\langle\bar\psi \psi\rangle$ as a 
function of temperature $T$, normalized to the vacuum quark condensate,
from lattice calculations with dynamical quarks. The condensate drops
rapidly to (almost) zero at the critical temperature $T_c$. 
(From \protect\cite{Boy95})}
\label{qq-cond}
\end{figure}

\begin{figure}
\centerline{\psfig{figure=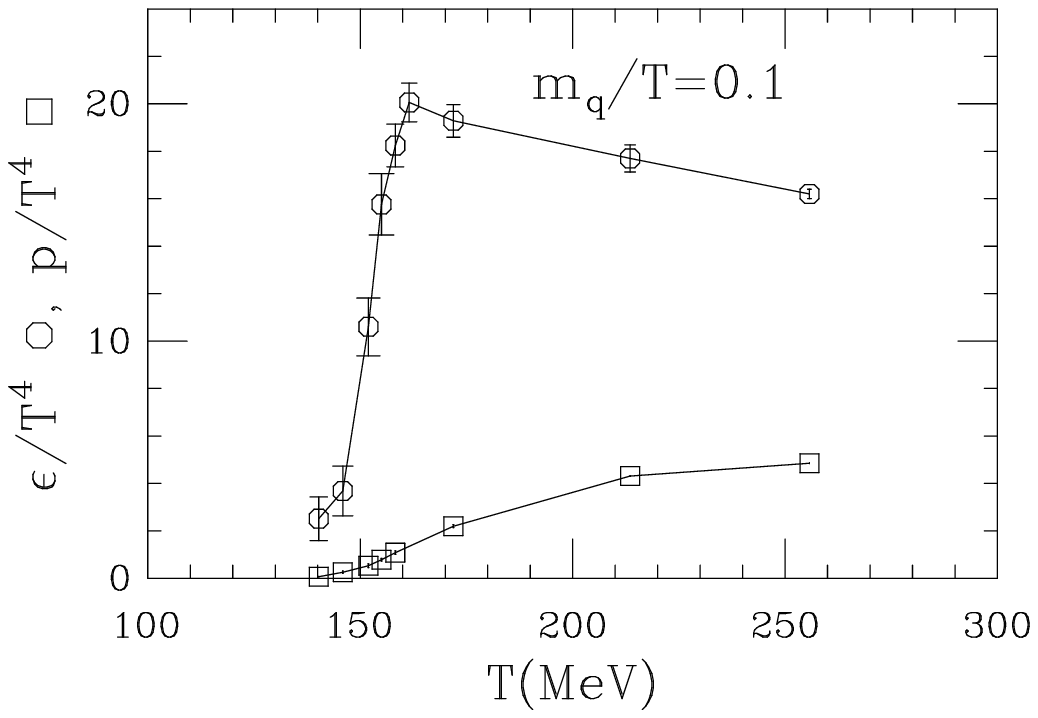,width=4in,angle=0}}
{\small Figure 2: Energy density $\epsilon$ (upper curve) and pressure 
$p$ (lower curve) from a numerical evaluation of QCD ``on the lattice'' 
with two light flavors of quarks. $\epsilon$ and $p$ are divided by $T^4$
to exhibit the sudden rise in the number of thermally excited degrees
of freedom at the critical temperature $T_c \approx 150$ MeV due to
liberation of color and chiral symmetry restoration. 
(From \protect\cite{Blu95})}
\label{lat-eos}
\end{figure}

\begin{figure}
\centerline{\psfig{figure=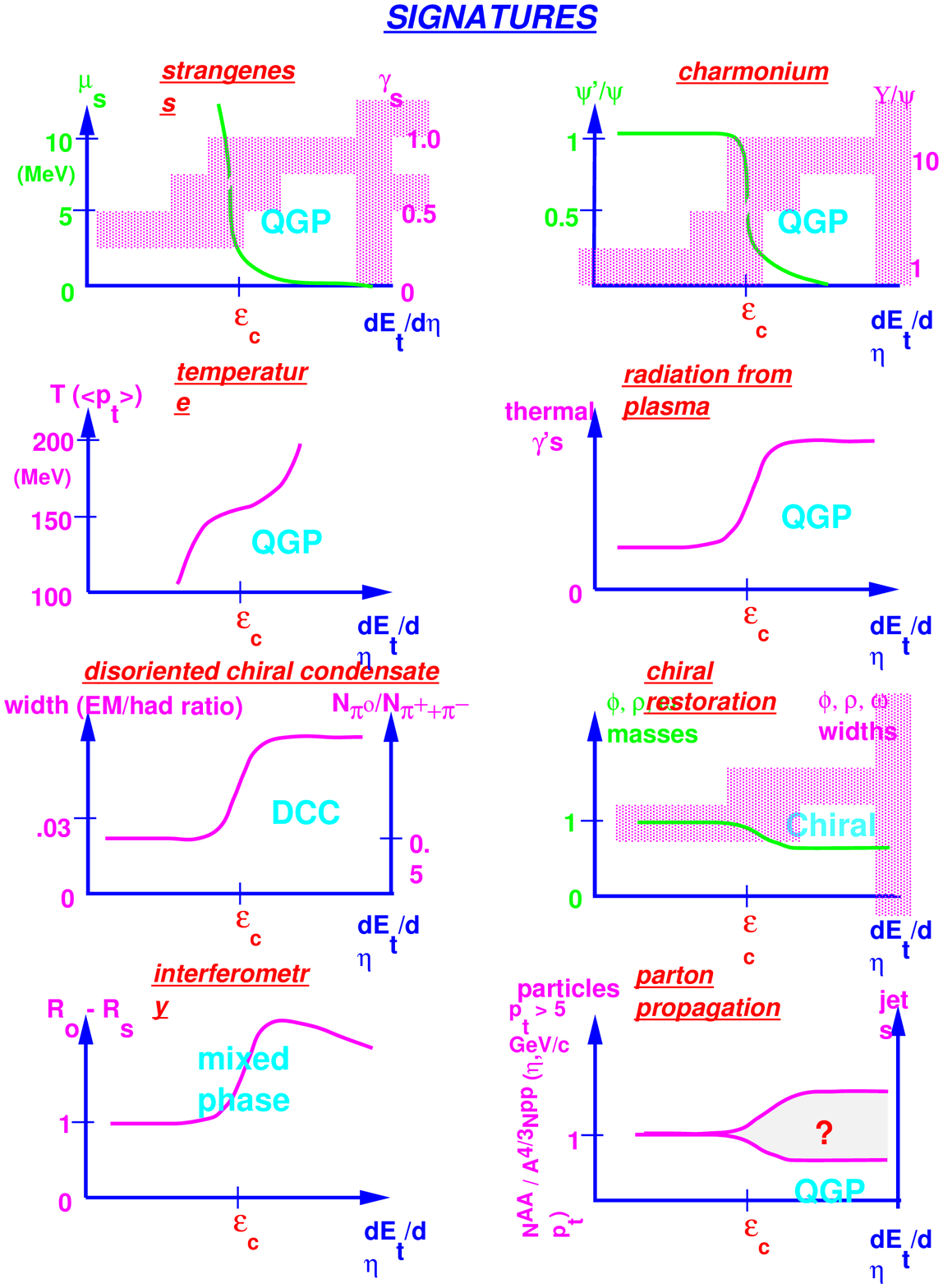,height=6in}}
{\small Figure 3: Signatures of quark-gluon plasma formation and 
the chiral phase transition. The expected behavior 
of the various signatures is plotted as a function of the 
measured transverse energy, which is a measure of the energy density,
in the region around the critical energy density $\epsilon_c$
of the transition. When two curves are drawn, the hatched curve
corresponds to the variable described by the hatched ordinate
on the right. See text of review for details.}
\label{signatures}
\end{figure}
\newpage

\begin{figure}
\centerline{\psfig{figure=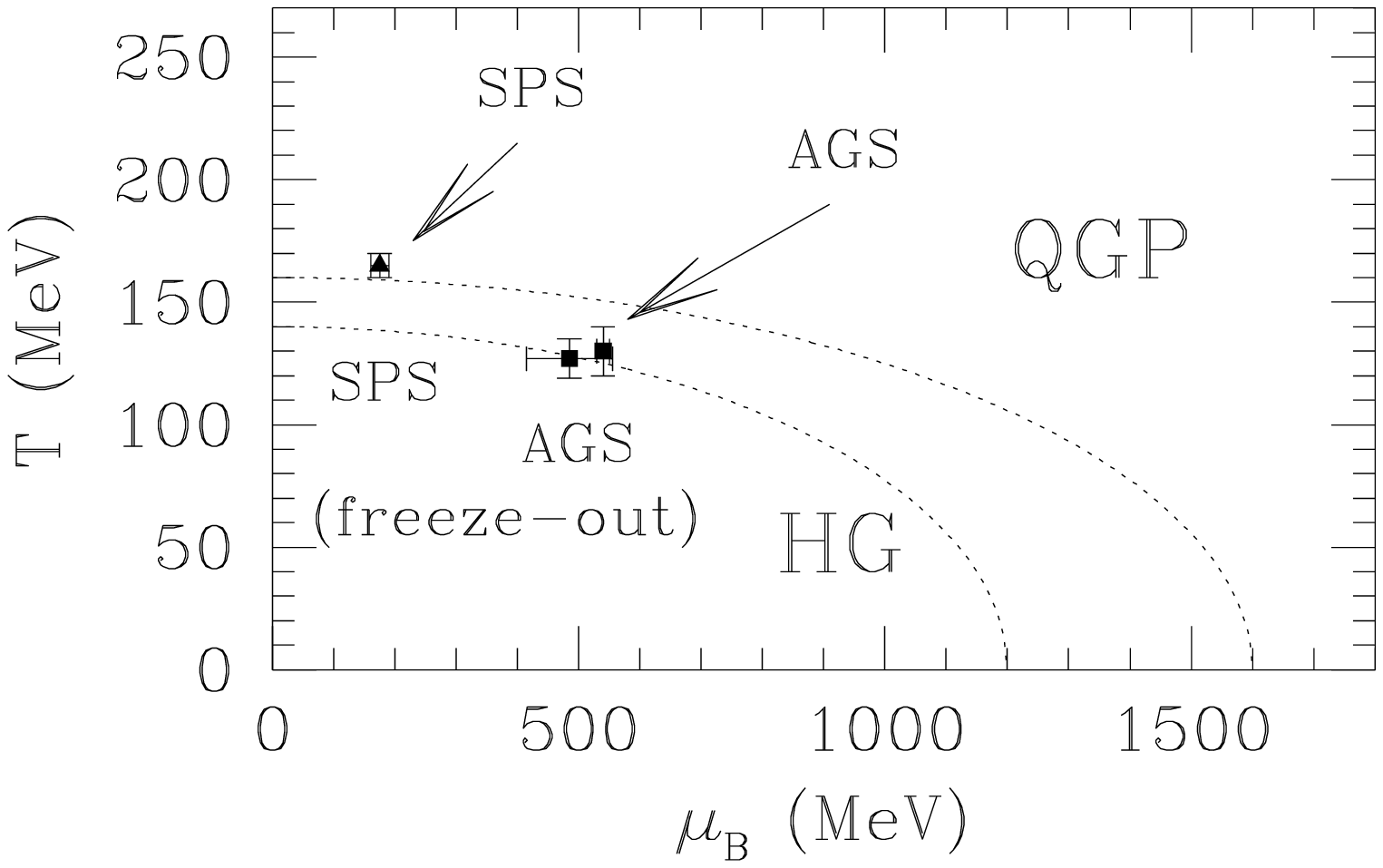,width=5in}}
{\small Figure 4: Thermal freeze-out parameters shown in the phase
diagram of nuclear matter.  The two dashed lines indicate location of
the expected phase boundary and its degree of uncertainty.  
The solid points with error bars show the freeze-out values
deduced from AGS and SPS data with flow; the arrows indicate how the
freeze-out conditions may be approached during the expansion of 
the fireball. The horizontal axis shows the baryon chemical potential 
$\mu$, which is a measure of the baryon density.}
\label{phase-data}
\end{figure}

\end{document}